\documentclass[10pt,psfig]{iopart}   
\usepackage{iopams}
\usepackage[dvips]{color}
\usepackage{graphicx}
\usepackage{bm}
\begin{document}
\title{
Evidence of triple collision dynamics in partial 
photo-ionisation cross sections of helium 
}

\author{Gregor Tanner}
\address{School of Mathematical Sciences,
University of Nottingham, University Park, Nottingham NG7 2RD, UK.}
\author{Nark Nyul Choi, Min-Ho Lee}
\address{School of Natural Science, Kumoh National Institute
of Technology, Kumi, Kyungbook 730-701, Korea.}
\author{Achim Czasch, Reinhard D\"orner}
\address{Institut f\"ur Kernphysik, Goethe-University Frankfurt,
Max-von-Laue-Str.\ 1 60438 Frankfurt, Germany
}

\date{\today}

\begin{abstract}
Experimental results on {\em partial} photo-ionisation cross sections 
of helium are analysed in the light of recent advances in the semiclassical 
theory of two-electron atoms. 
Byun {\em et al} \cite{BCLT06} predict that the {\em total} 
photo-ionisation cross section below the double-ionisation threshold 
can, semiclassically, be described in terms of contributions associated
with classical orbits starting and ending in the triple collision. 
The necessary modifications of the semiclassical theory for 
partial cross sections is developed here.
It is argued that partial cross sections are also dominated  
by the triple collision dynamics.
The expected semiclassical contributions can be identified in the Fourier 
transformation of the experimental data. 
This clearly demonstrates for the first time 
the validity of the basic assumptions made in \cite{BCLT06}.  
Our findings explain furthermore in a natural way the self-similar 
structures observed in cross section signals for different channel numbers.     
\end{abstract}

\maketitle
 
The rich resonance spectrum of two-electron atoms below the double
ionisation threshold has been explored both experimentally and 
numerically up to principle quantum numbers $N \approx 13 - 17$ 
of the remaining one-electron atom after ionisation
\cite{Pue01,Jia04,Cza04,Cza05,Jia06,Del06,Bue95, HG02, LMTL05}. 
Progress towards even higher $N$ values thus moving closer to the
three-particle break-up threshold $E=0$ are hampered experimentally by 
the limited photon energy resolution and numerically 
by the high-dimensionality of the system. 
The complexity of the classical dynamics and the large number of 
degrees of freedom of the system have so far also restricted semiclassical 
calculations of individual resonances to subsets of the full spectrum and 
again small $N$ values \cite{TRR00,CLT04}. Until recently it was thought
that similar restrictions also apply to a semiclassical treatment of total 
and partial photo-ionisation cross sections for $E<0$. The resonance density 
increases dramatically for energies approaching the double-ionisation 
threshold and individual resonances overlap and interfere leading to a 
strongly fluctuating cross section signal which decreases in amplitude 
towards the threshold \cite{Cza04,Cza05, Jia06}. The strong interaction 
between resonance poles can be regarded as a signature of the 
underlying chaotic classical dynamics which is also reflected in the 
resonance spacing distribution which shows a gradual transition towards 
that of random matrix theory approaching the double ionisation threshold 
from below \cite{Pue01,Jia06,LMTL05}.

We will consider in the following partial photo-ionisation cross sections of 
two-electron atoms in the asymptotic regime $E \to 0_-$. Near the threshold, 
electron-electron correlation effects dominate which can be observed 
directly in scaling laws such as Wannier's celebrated threshold law for 
double ionisation \cite{Wa53,KSA88} or in slow electron ionisation 
experiments taken across the threshold \cite{exp-zero}; 
a cusp-like structure in this ''zero-kinetic energy'' ionisation cross 
section is observed which has been interpreted in terms of classical escape 
along the Wannier ridge leading again to a threshold law with Wannier's 
exponent both below and above $E=0$
\cite{theo-zero}. Zero-kinetic energy spectroscopy below the
double ionisation threshold does, however, not resolve the complex resonance
structure of the three-particle compound and the experimental signal
contains thus little information about the mostly chaotic scattering dynamics
of the underlying classical three-body Coulomb system.

Semiclassically, total photo-ionisation cross sections can be described in 
terms of closed orbit theory (COT). The theory was developed for systems 
such as hydrogen in external fields \cite{Bog88,DD88} or in the context of
quantum defect theory for many electron atoms \cite{Del94,QD,COTS} for
which the dynamics near the origin is regular. Recently, the necessary 
modifications for a closed orbit treatment of the total photo-ionisation 
cross section for two-electron atoms have been 
presented in \cite{BCLT06}. Starting point of a COT is the total
cross section in dipole approximation written in the form
\begin{equation} \label{cs-G}
\sigma(E) = -4 \pi \, \alpha \, \hbar \omega \,
\Im \langle D \phi_i |G(E) |D \phi_i\rangle
\end{equation}
where $\phi_i$ is the wave function of the initial bound state,
$D = {\bm \pi} \cdot {\bf r}$ is the dipole operator,
$\bm \pi$ is the polarisation of the incoming photon with 
angular frequency $\omega$, and $G(E)$ is the Green function of the 
system at energy $E = E_i + \hbar \omega$; furthermore $\alpha = e^2/\hbar c$
is the fine-structure constant. In the semiclassical limit, the support of 
the wave function $\phi_i$ shrinks to zero relative to the size of the system 
reducing the integration in (\ref{cs-G}) to an evaluation of the Green 
function at the origin in the limit $E\to 0_-$. Writing the Green function 
in semiclassical approximation \cite{Bog88,DD88,Gut90} thus leads to a 
summation over contributions from classical trajectories starting and ending 
at the origin.

For two-electron atoms, the origin $ {\bf r} = ({\bf r}_1, {\bf r}_2) = 
{\bf 0}$ represents the point where both electrons reach the nucleus 
simultaneously, that is, all three particles collide. 
Note that we work in the infinite nucleus mass approximation, that is, the 
position of the nucleus is fixed at the origin. The presence of such triple 
collisions demands a careful re-evaluation of COT in the light of the 
three-body dynamics near the origin. The triple-collision itself forms 
a non-regularisable singularity of the classical equations of motion, that is, 
trajectories ending in a triple collisions can not be continued through the 
singularity. Trajectories coming close to a triple collision become 
extremely sensitive to initial conditions and nearby orbits approaching 
the collision point can be scattered into arbitrarily large angles. 
The triple collision singularity is in that sense infinitely unstable.
This is in contrast to binary collisions which can be regularised
by a suitable space and time transformation such as described in \cite{RTW93}
leading to a smooth phase space flow in the vicinity of the collision.  

A semiclassical treatment of photo-ionisation starting from (\ref{cs-G})
needs to take into account classical trajectories beginning and ending near 
the three-body collision $R = 0$ where $R = ({\bf r}_1^2 + {\bf r}_2^2)^{1/2}$
is the hyper-radius; the set of closed orbits usually employed in COT, namely 
those emerging out of and returning exactly to the triple collision point 
$R=0$, are infinitely unstable and give a vanishing contribution to 
semiclassical expressions. These {\em closed triple collision 
orbits} (CTCO) act, however, as guiding centres for phase space regions
leaving and returning to the triple-collision region. Taking the semiclassical 
limit $E \to 0_-$ is equivalent to $R_0\to 0$  in appropriately rescaled 
coordinates where $R_0$ characterises the size of the initial wave function 
$\phi_i$. By considering these limits carefully, it has been shown in 
\cite{BCLT06} that the amplitude of the fluctuations in the total 
photo-ionisation cross section decays with a power-law according to 
$\sigma_{fl} \propto |E|^{\mu}$ with predicted exponent 
\begin{equation} \label{mu}
\mu= \frac{1}{4}\Re \left[\sqrt{\frac{100 Z-9}{4Z-1}} +
2\sqrt{\frac{4 Z - 9}{4Z -1}}\right]\, .
\end{equation}
with $Z$, the charge of the nucleus.
One obtains, for example, $\mu = 1.30589...$ for helium
with $Z=2$. The exponent can be obtained as a combination of 
stability exponents of the triple-collision singularity and differs from 
Wannier's exponent which describes double ionisation processes. The Fourier
components of the fluctuations can furthermore be associated directly with 
CTCO's. This behaviour has been confirmed by quantum 
calculations in collinear helium, the restricted three-body Coulomb problem 
where the dynamics takes place along a common axis \cite{BCLT06}.

\begin{figure}
\begin{minipage}[c]{0.90\textwidth}
\begin{center}
\includegraphics[scale=.60]{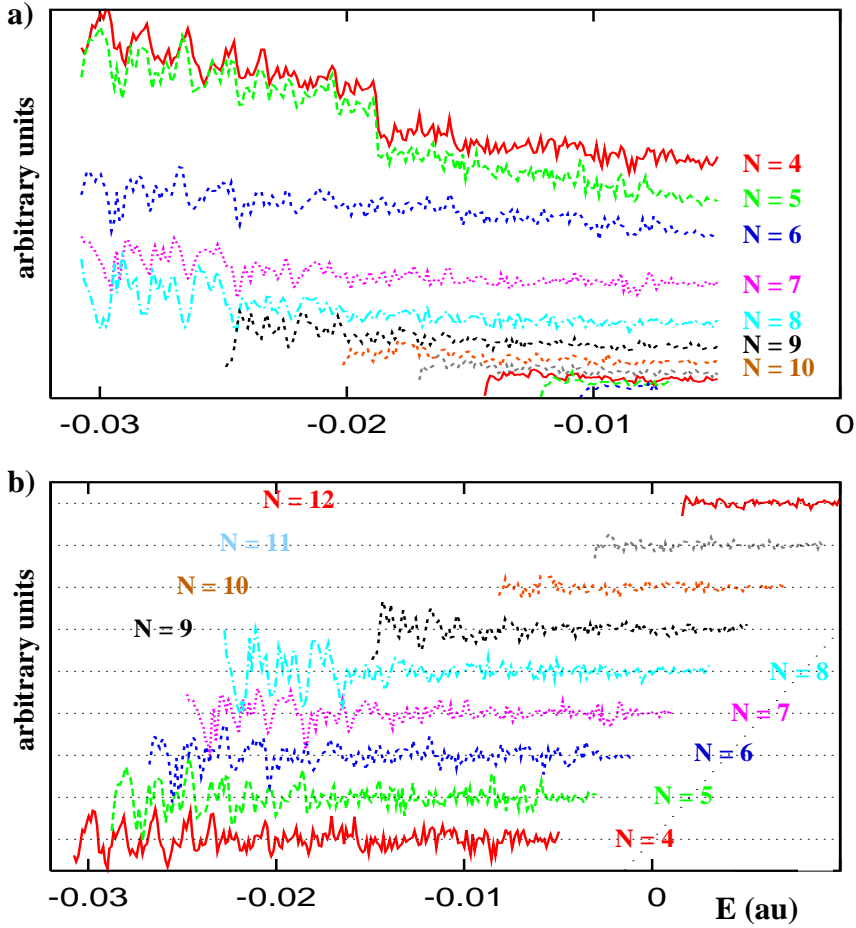}
\caption[fig:fig1]{a) Partial cross sections as measured in \cite{Cza04,Cza05};
(the jump in the signal for $N=4,5$ at $E \approx -0.02$ au is due to 
experimental reasons; the signal for $N=4$ was in addition shifted down 
towards $N=5$). b) The fluctuating part of the cross sections for different 
channel numbers obtained from a) after subtracting a smooth background 
contribution.
}
\label{fig:fig1}
\end{center}
\end{minipage}
\end{figure}

In this letter, we analyse the experimental data on partial photo-ionisation 
cross sections for helium presented in \cite{Cza04,Cza05} (and shown in  
Fig.\ \ref{fig:fig1}a), in the light of the theoretical predictions 
\cite{BCLT06}. The experiment was conducted using synchrotron radiation 
with a resolution of 4 meV revealing partial cross sections up 
to energies 78.85 eV above the ground state and reaching ionisation 
channels of the order $N  \approx 13$. The kinetic energy of the 
outgoing electron was measured which yields information on the state of the 
remaining ${\rm He}^+$ - ion; for experimental details, see \cite{Cza04}.  
Ref.\ \cite{Cza05} notes in particular, that the cross sections $\sigma_N$ 
for different channels show similar patterns up to an overall 
shift in energy, see for example the data for $N=4$ and $N=5$ in the region 
$E < -0.02$ au in Fig.\ \ref{fig:fig1}a; (in Fig.\ \ref{fig:fig1} and 
throughout the paper, atomic units (au) are employed). The phenomenon could 
be reproduced in R-matrix calculations published in the same article. 
In Fig.\ \ref{fig:fig1}b, the fluctuating part of the experimental partial 
cross sections are shown after subtracting numerically a smooth background 
contribution.  

In the following, we 
will argue semiclassically that CTCOs also dominate the fluctuations in 
partial ionisation cross-sections and show that collision orbits can 
actually be detected in the experimental data. This leads naturally 
to a semiclassical explanation 
for the similarities seen in the cross-sections $\sigma_N$ for 
different $N$. The experimentally observed slight shifts in energy 
between the patterns in different partial cross sections can then
be explained in terms of phase differences between direct and 
indirect contributions as discussed in more detail later. 
The experimental data clearly show a decrease in the amplitude of 
the fluctuations as $E\to 0_-$; the exponents characterising the mean 
power-law decay of the fluctuations can at present not be extracted from 
the data with sufficient accuracy to allow a comparison with theoretical 
predictions; they will thus not be considered here. A detailed 
theory of threshold laws for partial cross sections will be presented 
elsewhere \cite{CLT07}.

We start by expressing the partial cross sections $\sigma_N$
in terms of the retarded Green function $G(E)$,
from which a semiclassical theory can be developed.
The outgoing solution of the inhomogeneous Schr\"odinger equation
\begin{equation} \label{SE}
(E - H) |\chi> = D |\phi_i>
\end{equation}
is written as
\begin{equation} \label{chi}
|\chi> = G(E)  |\Phi_i> ~~ \mbox{with} ~ |\Phi_i> = D |\phi_i> ,
\end{equation}
and using the notation introduced after Eq.\ (\ref{cs-G}).
In the asymptotic limit $R \to \infty$ which corresponds for 
fixed $E < 0$ to either $r_1 ~ \mbox{or} ~ r_2 \to \infty$,
the Schr\"odinger equation (\ref{SE}) is homogeneous and separable. 
One obtains asymptotically for the inhomogeneous solution  $|\chi>$  
in the ionisation channel $(N \lambda)$
\begin{equation} \label{asym}
<r_1 N \lambda  | \chi > = <r_1 N \lambda | G(E) |\Phi_i> \propto
f_N^+(r_1)
~~~ \mbox{for} ~~ r_1 \to \infty 
\end{equation}
where $\lambda =\{ L M l_1 l_2 \}$ denotes the quantum numbers describing 
angular momenta and $N$ is the principal quantum number of the remaining ion;
furthermore, $f_N^+ (r_1)$ is the outgoing Coulomb function
\begin{equation}
f_N^+ (r_1) = \exp [ i (k_N r_1 + \frac{Z-1}{k_N}\ln r_1)]
\end{equation}
with $k_N = \sqrt{2 (E - E_N)}$, the momentum of the outgoing electron. 
After writing the cross section as flux through the surface $R = const$, 
taking the limit $R \to \infty$, and using the asymptotic form of the 
inhomogeneous solution (\ref{chi}) as given in (\ref{asym}), we express
the partial cross section $\sigma_N$ in the form
\cite{Fri05}
\begin{equation} \label{sigmaN}
\sigma_N = \sum_\lambda \sigma_{N \lambda}
= \sum_\lambda 2 \pi \alpha \omega k_N
\lim_{r_1 \to \infty}  |G(r_1 N \lambda; E)|^2
\end{equation}
where $G(r_1 N \lambda; E) =  <r_1 N \lambda | G(E) |\Phi_i>$ 
and $\sigma_N$ becomes independent of $r_1$.

The matrix elements $G(r_1 N \lambda; E)$
of the Green function can in semiclassical approximation be described in terms
of classical trajectories starting at $R \le R_0$ near the origin 
and reaching $r_1 \to \infty$ with fixed energy $E_N$ of the hydrogen like atom
and fixed quantum numbers $\lambda$, thus determining the kinetic 
energy $\epsilon_N = k_N^2/2$ of the escaping electron. For details see
\cite{CLT07}; a similar treatment applied to transport in quantum wires with 
fixed channel numbers can be found in \cite{JBS90}. We will distinguish direct 
escape leading from the initial region $R < R_0$ directly to ionisation and 
indirect contributions from trajectories entering into a predominately 
chaotic phase space region before ionisation, that is, 
\begin{equation}
G(r_1 N \lambda; E) =G_{dir}(r_1 N \lambda; E) + G_{ind}(r_1 N \lambda;E) .
\end{equation}

The cross section $\sigma_{N \lambda}$ can now be written in terms 
of a  smooth background part and a fluctuating contribution according to
$\sigma_{N \lambda}
= \sigma_{N \lambda}^0 + \sigma_{N \lambda}^{(fl)}$
where
\begin{eqnarray}
\sigma_{N \lambda}^0 &=& 2\pi \alpha \omega k_N
|G_{dir}(r_1 N \lambda; E)|^2\label{sigmadir}\\
\sigma_{N \lambda}^{(fl)} &=& 4\pi \alpha \omega k_N \,
\Re \left[G_{dir}^*(r_1 N \lambda; E) G_{indir}(r_1 N \lambda; E) \right]. 
\label{sigmafl}
\end{eqnarray}
In (\ref{sigmafl}), the term $|G_{indir}|^2$ is neglected as it contains 
contributions from pairs of indirect trajectories which give rise to
lower order corrections in the asymptotic limit $|E/E_N| << 1$.

\begin{figure}
\begin{minipage}[c]{0.90\textwidth}
\begin{center}
\includegraphics[scale=.50]{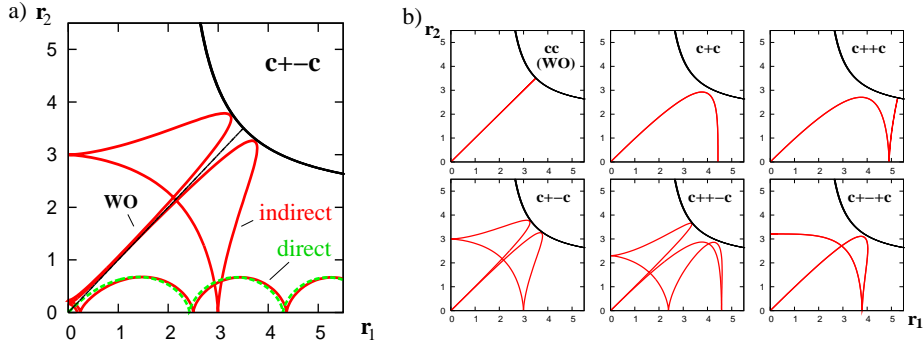}
\caption[]{\small Triple collision dynamics and closed triple collision orbits: 
a) trajectories starting near the triple collision point $r_1 = r_2 = 0$ and 
leading to ionisation with fixed energy of the escaping electron; 
dashed (green) line: direct path; thick full (red) line: indirect path, here 
close to the CTCO $c+-c$; b) some short CTCOs together with binary symbol code. 
}
\label{fig:fig2}
\end{center}
\end{minipage}
\end{figure}

We turn now to a brief discussion of the classical dynamics in 
two-electron atoms. By introducing the scaling transformation \cite{RTW93}
\begin{equation}
{\bf r} = \tilde{\bf r}/|E|;
\quad {\bf p} = \sqrt{|E|}\tilde{\bf p};
\quad S = \tilde{S}/\sqrt{|E|},
\quad {\bf L} = \tilde{\bf L}/\sqrt{|E|},
\end{equation}
one studies the dynamics at fixed energy $E=-1$ where 
$\tilde{\bf r}, \tilde{\bf p}$ denote the scaled coordinates and momentum and 
$\tilde{\bf L}$ is the total scaled angular momentum.  The region $R \le R_0$ 
containing the initial state $\phi_i$ shrinks according to 
$\tilde{R}_0 = |E| R_0 \to 0$ for $E\to 0_-$, that is, the initial conditions 
approach the triple collision. Likewise, the kinetic energy of the outgoing 
electron diverges according to $\tilde{\epsilon}_N = \epsilon_N/|E|$ 
for fixed $N$.  Expressing the angular momentum in scaled coordinates,
we have $\tilde{\bf L} \to 0$ as $E\to 0$ for fixed $\bf L$; the classical 
dynamics takes place closer and closer to the zero angular momentum manifold. 
It can therefore asymptotically be characterised by trajectories in the 
invariant subpace $\tilde{\bf L} = 0$ which has only three degrees of 
freedom \cite{CLT04}. The part of the dynamics contributing to the 
semiclassical Green function $ G = G_{dir} + G_{ind}$ is in scaled coordinates 
formed by trajectories starting closer and closer to the triple collision $R =0$ as
$|E| \propto \tilde{R}_0 \to 0$. The triple collision singularity $R=0$ itself
 has a non-trivial structure and is equivalent to the phase space dynamics at 
$E=0$. We will here describe only those features of the dynamics near $R=0$ 
which are important for understanding the semiclassical approximations; for 
more details see \cite{CLT04, LCT05,BGY98}. Most initial conditions near the 
triple collision will give rise to trajectories leading to immediate 
ionisation of one electron. Direct orbits escaping with  
kinetic energy $\epsilon_N$ will contribute to the direct escape term 
$G_{dir}$; an example of such a trajectory (dashed line) is shown in 
Fig.\ \ref{fig:fig2}a. Only a fraction of the phase space near $R=0$ 
can enter a chaotic scattering region; these orbits all move 
out along the so-called Wannier orbit (WO), the trajectory of 
collinear and symmetric electron dynamics with $r_1 = r_2$. 
An example of such a indirect orbit is shown in Fig.\ \ref{fig:fig2}a 
(thick full line).  Likewise, trajectories inside the chaotic phase 
space can only leave this region after approaching the triple collision 
along the WO. Escaping from the triple-collision region towards ionisation with 
asymptotic kinetic energy $\tilde{\epsilon}_N = \epsilon_N/|E| \to \infty$ 
as $E\to 0_-$ can be achieved only by coming closer and closer to the triple 
collision.  The point of closest approach, $\tilde{R}_N$, of trajectories 
escaping with kinetic energy $\tilde{\epsilon}_N$ vanishes for fixed $N$ like 
$\tilde{R}_N \propto |E|$.  The indirect part of the Green function 
$G_{ind}(N)$ is thus semiclassically described in terms of orbits starting 
and ending (in scaled coordinates) closer and closer to the triple collision 
before escaping with  kinetic energy $\tilde{\epsilon}_N$. These trajectories
approach a proper CTCO asymptotically. The phase space regions contributing 
to the partial Green
function $G_{indir}(N)$ can thus  be characterised in terms of CTCO's in 
the same way as the total cross section \cite{BCLT06}. 
Triple collision orbits only occur in
the so-called eZe space \cite{CLT04}, a collinear subspace of the full three
body dynamics where the two electrons are on opposite sides of the nucleus
\cite{RTW93}. As $\tilde{R}_0, \tilde{R}_N \to 0$, only orbits coming close 
to the eZe space can start and return to the triple collision and they will 
do so in the vicinity of a CTCO.  The dynamics in the eZe space
is relatively simple as it is conjectured to be fully chaotic
with a complete binary symbolic dynamics. The symbolic coding of a trajectory 
is here defined as:\\[.1cm]
\indent
$-$ : electron 1 collides with the nucleus - $r_1 = 0$;

$+$ : electron 2 collides with the nucleus - $r_2 = 0$;

$c \;\,$ : triple - collision.\\[.1cm]
The completeness of the symbolic dynamics implies that there is exactly one
CTCO for every finite binary symbols string; the shortest is the 
WO with code $cc$. In Fig.\ \ref{fig:fig2}b, some short CTCOs are shown; 
note that orbits whose symbol code is related by the operation 
$+ \leftrightarrow -$ are mapped onto each other by the particle exchange 
symmetry $r_1 \leftrightarrow r_2$ and are thus equivalent. 

CTCOs have been shown to be important for the total cross section
\cite{BCLT06} giving the main contributions in a modified semiclassical
closed orbit treatment of photo-ionisation. The fact that CTCOs also
enter partial cross section here is due to the above-mentioned escape mechanism 
which is special to two-electron atoms.  Swarms of
indirect trajectories starting at $R = R_0$, following a similar path in the 
chaotic region and then escaping with fixed kinetic energy $\epsilon_N$ 
can for $E \to 0$ be linked to a single CTCO together with a 
direct-escape orbit. Likewise, the phase space region leading from 
$R<R_0$ to direct escape can be represented by a single trajectory. We 
thus write
\begin{eqnarray}
G_{dir}&\approx& A_0(N \lambda)
e^{i S_0(r_1 N \lambda; E) - i \nu_0(r_1 N \lambda; E) \pi/2} 
\, ;\label{Gdir} \\
G_{ind}&\approx& 
e^{i S_0(r_1 N \lambda; E) - i \nu_0(r_1 N \lambda; E) \pi/2} \nonumber \\
& & \times \sum_{\mbox{CTCO}} A_j(N \lambda; E)
e^{i S_j(E) - i \nu_j \pi/2}  \label{Gind} .
\end{eqnarray}
Here $S_0$, $\nu_0$ are the action and Maslov index along the direct orbit, 
respectively, and $S_j$, $\nu_j$ are those along a CTCO starting from and 
ending in the triple collision.  \footnote{We note that extra phase 
contributions arise due to the stationary phase approximation leading to the 
fixed kinetic energy condition in the outgoing channel;
these phases are not energy dependent and are here absorbed in the 
pre-factors.} The initial wave function $\Phi_i$ and the dynamics in phase 
space leading to direct orbits is not sensitive to the threshold energy 
$E =0$. The pre-factor $A_0$ will thus pick up a smooth energy dependence and 
can be treated as constant at the threshold. The singular behaviour of the 
classical dynamics near the triple collision for CTCO contributions leads to 
a universal scaling behaviour of the terms  $A_j$ as $E \to 0_-$ as 
described in \cite{BCLT06}; the semiclassical analysis suggests the 
same scaling law as for the total cross section, that is, 
\begin{equation} \label{Aj}
A_j(N\lambda; E) = a_j(N\lambda) \, |E|^\mu
\end{equation}
for $E \to 0_-$ and fixed $N$ with  exponent given by Eq.\ (\ref{mu}).
From (\ref{sigmaN}), (\ref{sigmafl}), (\ref{Gdir}), (\ref{Gind}) and (\ref{Aj}),
we obtain an approximation to the fluctuating part of
the partial cross sections in the form
\begin{equation} \label{PPIC}
\sigma_{N}^{(fl)} \approx 4\pi \alpha \omega k_N |E|^\mu \,
\Re \left( \sum_{\mbox{CTCO}}
{\cal A}_j(N)
e^{i z \tilde{S}_j - i \nu_j \pi/2} \right)
\end{equation}
where the pre-factor ${\cal A}_j(N)$ are of the form ${\cal A}_j(N) =
\sum_\lambda a_j(N\lambda) A_0(N\lambda)$ and $z = 1/\hbar \sqrt{|E|}$;
note, that the ${\cal A}_j(N)$ are in general complex valued and will 
add extra (energy independent) phases.

\begin{figure}
\begin{minipage}[c]{0.90\textwidth}
\begin{center}
\includegraphics[scale=.60]{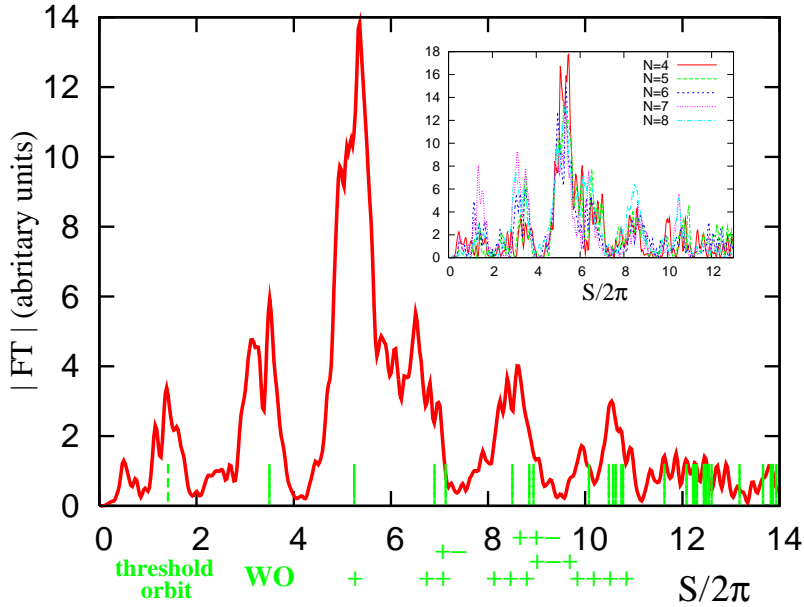}
\caption[]{\small The Fourier spectrum of the fluctuating part of the 
partial cross section in Fig.\ \ref{fig:fig1} averaged over 
$N = 4 - 8$; the actions of CTCOs are marked on the axis together with
their respective binary symbol code.  Inset: individual Fourier signals.
}
\label{fig:fig3}
\end{center}
\end{minipage}
\end{figure}

The semiclassical treatment above suggests that a Fourier transformation of 
$\sigma_N^{(fl)}$ in terms of the variable $z$ would reveal peaks at 
the actions of classical CTCOs. Such an analysis has been carried out 
using the experimental data from \cite{Cza04,Cza05} setting the origin
of the energy scale at the double ionisation threshold, that is, 79.014 eV 
above the ground state of helium. After subtracting a smooth background
by fitting a low-order polynomial to the experimental data, the fluctuating
signal as shown in Fig.\ \ref{fig:fig1}b is transformed using the Lomb - 
algorithm \cite{NR94}, a Fourier transform technique particularly useful for 
unevenly spaced data sets.  (Note that the experimental data are taken at 
uniform steps in $E$ which leads to an increasing step-size in $z$.) The peak 
heights in the experimental partial cross sections are roughly of the same 
size for fixed energy independent of $N$. The corresponding Fourier signals 
thus show structures of similar size, see inset in Fig.\ \ref{fig:fig3}; to 
enhance the resolution, we averaged over the transformed cross sections for 
$N=4 - 8$ for which data sets in energy intervals of the same length 
are available. The resulting averaged Fourier signal is shown in Fig.\
\ref{fig:fig3} together with the values of the actions of short CTCOs 
such as depicted in Fig.\ \ref{fig:fig2}b.

Despite the fact that the energy range available is rather limited, there are
clear correlations between the positions of the pronounced peaks and the 
actions of CTCOs; 
(note, that the high-resolution Fourier signal presented in \cite{BCLT06}
for the total cross section was obtained using a much larger energy range - up 
to $N = 50$ - in the collinear eZe model). The Fourier signal shown in 
Fig.\ \ref{fig:fig3} is a clear indication that photo-ionisation cross sections
of three dimensional helium are indeed dominated by the triple
collision dynamics in the collinear eZe subspace. The similarities in the 
partial cross sections noted in \cite{Cza05}, are 
now naturally explained in terms of the semiclassical expansion (\ref{PPIC});
the dominant terms contribute asymptotically with the same phase - the 
actions of CTCOs - which will produce similar modulations in the overall 
signal for different $N$. The slight energy shifts observed in the cross 
section pattern when comparing signals for nearby $N$ values as can 
be seen Fig.\ \ref{fig:fig1} are caused by energy independent phases in the 
$A_j(N)$'s entering Eq.\ (\ref{PPIC}); a detailed analysis will be presented 
in  \cite{CLT07}.

Due to the degeneracy of the $c+c$ and $c-c$ orbits related by particle 
exchange symmetry, the peak associated with the code $+$ is about twice 
as high as the WO - peak. A similar phenomena was observed for the total 
cross section in \cite{BCLT06}. Otherwise, the semiclassical amplitudes 
decrease in general exponentially with the length of the symbol code 
which can also be seen in the Fourier data. Furthermore, a peak at an 
action smaller than that of the shortest CTCO, the WO with action 
$\tilde{S}/2 \pi = 3.5$, can be observed in Fig.\ \ref{fig:fig3}. We note, 
that the peak is at $\tilde{S}/2 \pi = \sqrt{2}$, which is the action of the 
asymptotic threshold orbit where the outer electron escapes with zero 
kinetic energy leaving the inner electron at a total energy 
$\tilde{E}_N = -1$. The peak may thus stem from non-perfect cancellations 
between actions for direct and indirect contributions especially near 
the channel thresholds. It is a feature special to partial cross sections 
and is not expected to be observed in total photo-ionisation signals.\\

In conclusion, we show that the fluctuations in the partial photo-ionisation 
cross section in helium below the double ionisation threshold are 
semiclassically dominated by contributions from closed triple collision 
orbits. This explains naturally the similarities observed in different
partial cross sections and is a clear evidence that both partial and 
total cross sections are dominated by the low-dimensional collinear 
eZe dynamics \cite{BCLT06}. It is at present not possible to extract
threshold laws such as stated in Eq.\ (\ref{Aj}) from the data; a detailed 
account of the theory including the $N$-dependence of the amplitude
terms due to the direct and indirect contributions will be presented 
elsewhere \cite{CLT07}.

\ack
Nark Nyul Choi acknowledges financial supports by the
Kumoh National Institute of Technology for the project
2006-104-067 and the Royal Society for visiting the University
of Nottingham where parts of this work were carried out.\\

\end{document}